\documentclass[11pt]{article}
\usepackage[paper=a4paper,dvips,top=2.5cm,left=1cm,right=1cm,bottom=2.6cm,footskip=2.2cm,voffset=0.0cm, marginparwidth=18mm]{geometry}
\usepackage{amsfonts,amsmath,amsthm,amssymb}
\usepackage[svgnames]{xcolor}
\usepackage{fix-cm}
\usepackage{sectsty}
\usepackage{fancyhdr}
\usepackage{lastpage}
\usepackage{graphicx}
\usepackage{subfigure}
\usepackage{caption}
\usepackage{url}
\usepackage{enumerate}
\usepackage{paralist}
\usepackage{multicol}
\usepackage{color}  
\usepackage{float}  
\usepackage{epsfig}
\usepackage{subfigure}
\usepackage{hyperref}   
\hypersetup{colorlinks=true,linkcolor=beamer@PRD, citecolor=beamer@PRD}
\usepackage{authblk}  
\usepackage{cite}  
\usepackage{todonotes}
\usepackage{ulem} 

\renewcommand\eqref[1]{\textcolor{beamer@PRD}{(}\ref{#1}\textcolor{beamer@PRD}{)}}

\definecolor{beamer@PRD}{RGB}{46,48,146}

\begin{document}
\title{   { A Novel Application  of Quantum Speed Limit to String Theory } }
\author{\small{Arshid Shabir$^1$,  Salman Sajad Wani$^4$, Raja Nisar Ali$^3$, S. Kannan$^5$,
    Aasiya Sheikh$^6$, Mir Faizal$^{7,8}$, Javid A. Sheikh$^2$,  Seemin Rubab$^{1}$,  Saif Al-Kuwari $^4$ } \\
\textit{\small $^{1}$Department of Physics,  National Institute of Technology,
Srinagar  190006, India}
\\
\textit{\small $^2$Department of Physics, University of Kashmir, Srinagar  190006, India}
\\
\textit{\small $^{3}$
Department of Physics, Central University of Kashmir, Tulmulla Ganderbal  191131, India}
\\
\textit{\small $^{4}$Qatar Center for Quantum Computing, College of Science and Engineering, Hamad Bin Khalifa University, Qatar}
\\
\textit{\small $^{5}$ISRO Inertial Systems Unit, Thiruvananthapuram 695013,  India}
\\
\textit{\small $^6$Design and Manufacturing Technology Division, Raja Ramanna Centre for Advanced Technology, Indore 452013, India}
\\
\textit{\small $^{7}$Canadian Quantum Research Center, 204-3002, 32 Ave Vernon, BC V1T 2L7, Canada}
\\
\textit{\small{$^{8}$ Irving K. Barber School of Arts and Sciences, University of British Columbia Okanagan, Kelowna, BC V1V 1V7, Canada}}}
\date{}
\maketitle
\begin{abstract}
  In this work, we investigate the implications of the concept of quantum speed limit in string field theory.
  We adopt a novel approach to the problem of time on world-sheet based on Fisher information, and arrive at a  minimum time for a particle state to evolve into another particle state. This is done using both the Mandelstam-Tamm bound and the
 Margolus-Levitin bound.  This implies that any interaction has to be
  smeared over such an interval, and  any interaction in the effective  quantum field theory has to be non-local.  As non-local
  quantum field theories are known to be finite, it is expected that  divergences  should be removed from effective
  quantum field theories due to  the quantum speed limit of string theory. 
\end{abstract}	 
\addtolength{\footskip}{-0.2cm} 
 
It is known that divergences occur in quantum field  theories due to the point-like structure of the vertices, and are generally
removed by imposing a
cutoff \cite{1b}. However, even though this procedure works well for renormalizable field theories, it does not
work for non-renormalizable field theories. As many important field theories including gravity are not
renormalizable \cite{2b, 3b}, it s important to look at the idea of renormalization from a different perspective. For
any theory, renormalization is performed by imposing an ultraviolet cut-off, and this corresponds to choosing a maximum
energy scale in the theory. The method is not probed beyond that energy scale  however, due to the time-energy uncertainty principle, this would correspond to a finite time interval.
It is then possible to use  the interpretation of  the time-energy uncertainty principle for obtaining the quantum speed
limit in terms of variance, i.e.,  the interpretation used in Mandelstam-Tamm bound \cite{6}.
In this interpretation, the
quantity of time in the uncertainty principle is the minimum time taken by a quantum state to evolve into a different orthogonal
quantum state for a  system with a certain energy. This interpretation of the time-energy uncertainty principle is
also critical in the derivation of Lloyd's computation bound   \cite{4b}. Thus, considering this interpretation of the
uncertainty principle, there should be a minimum time for any particle state to evolve into any other particle state. This,
in turn, implies that it is impossible to have point-like vertices, and so divergences cannot occur. It
has been rigorously demonstrated that if the point interactions are replaced by non-local interactions, then the divergence
does not occur and the theory becomes finite  \cite{a1, a2, a3, a4, a5}. In this work, we analyze this aspect of
quantum field theories more rigorously using the quantum speed limit. Apart from  the Mandelstam-Tamm bound  \cite{6},
where the quantum speed limit is expressed in terms of variance of energy,  it is also possible to express it using
the mean energy in Margolus–Levitin bound \cite{7}.  

It is important to note that the quantum speed limit can be derived using  purely
geometric derivations   \cite{16,17,18,19,20}. 
The quantum speed limit has been used to investigate time-dependent systems \cite{10,11,12}, open quantum systems \cite{13,14,15} and various other quantum systems \cite{21,22,23,24,25,26,27,28}.
However, in this work, we investigate a novel application of quantum speed limit to properly analyze the minimum time required for a
particle to disintegrate into another particle. As the standard quantum field theories are constructed by assuming
that such interactions convert a certain particle into a different particle occur instantaneously, it would not be
possible to directly conduct such analysis using quantum field theories. Instead, we need a formalism in which the conversion of a
particle to a different particle can be dynamically analyzed. Interestingly, this can be done using the string theory, as in this
theory, the different particle states are just different quantum states of a single string. Thus, string theory can be
used to analyze the minimum time required for a string state (representing a certain particle) to evolve to a different string state
(representing a different particle). 

It is non-trivial to discuss time in string theory as time is not an observable in
quantum mechanics \cite{time12}. However, such  quantities, which are not associated with an observable, can be investigated
using Fisher information concept \cite{eq1, eq2, eq4, eq5, eq6, eq7}. It has been recently argued that time
in string theory can be defined as an unobservable variable and can only be indirectly probed $\hat{M}^2$ \cite{mass, mass1}.
The time probed by Hamiltonian does not give any information about the casual ordering of events on the world-sheet
due to time reparametrization invariance \cite{t12}.
In fact, due to the time reparametrization invariance, the world-sheet Hamiltonian does not contain any information, and cannot
be used to study  the dynamics of the system \cite{t12}. However, the $\hat{M}^2$ does contain Fisher information
and can be used to study the dynamics of the system \cite{mass, mass1}. 
We would like to clarify that by \emph{time} we mean an ordering parameter, which can be used to define the sequences
in which events occur. Such novel interpretations of time as an ordering parameter to define sequences of events
have been used in systems with time  reparametrization invariance \cite{pt12, pt14}. Such a novel approach to the problem of
time  has been used in the Wheeler-DeWitt approach \cite{wh, wh0}, loop quantum gravity \cite{wh2, wh21},
discrete quantum gravity \cite{wh4, wh41}, group field theory \cite{wh5, wh51}, quantized modified   gravity \cite{wh6, wh61},
and in the quantization of both brane world theories  \cite{wh7, wh71} and Kaluza–Klein geometries  \cite{wh8, wh81}. 
This is usually done using  a part of the Hamiltonian, to probe change and hence obtain a time variable.    The 
unobservable variable corresponding to this part of Hamiltonian   on the world-sheet is $\theta$, and hence it  can directly be related to time \cite{mass, mass1}. 
Here, we will use $\theta$ to discuss the evolution of world-sheet string states, as it can be
used to define evolution on world-sheet coordinates. 
We will represent this time on the world-sheet,  which corresponds  to the unobservable variable probed by
$\hat{M}^2$ (since it contains non-zero Fisher information), to clarify that it is not the unobservable
variable probed by the Hamiltonian. 

Thus, a minimal time would correspond to the optimum interval required for a particle state with a specific  $\hat{M}^2$
eigenvalue to evolve into a different particle state with a different $\hat{M}^2$ eigenvalue. As this time has a lower bound,
it is not possible to define point interactions, with the instantaneous conversion of a certain particle state to a
different particle state. This is what is indirectly done using renormalization by imposing
a cutoff \cite{1b}.  However, in this work, these results are obtained directly using the quantum speed limit and will be deduced
using both the Mandelstam-Tamm bound  \cite{6}  and the  Margolus–Levitin bound \cite{7} for the string theory. 

We will use the string coherent states to  analyze the quantum speed limit.
Since quantum speed limits offer broad restrictions on the speed of dynamical evolution, they are intimately related to
time-energy uncertainty relations. In this work, we will compute the quantum speed limit for the open string coherent state.
To do this, we take into account the evolution that is parameterized by $\theta$, which is the unobservable variable, probed by $\hat{M}^2$  operator. 
In   the light-cone gauge, we have  to consider the   string oscillations only in the $\{  X^i\}_{i=1}^{24} $  direction, with  $\{i\}$,  taking  values in $\{1,2,3,...,24\}$. Hence,  the string algebra in light-cone gauge  can be written as 
\begin{equation}
[\hat{\alpha}_m^i ,\hat{\alpha}_n^j] =m\eta^{ij} \delta_{m+n,0}.
\end{equation} 
We note that for $ n =  -m $, 
$
[\hat{\alpha}_m^i ,\hat{\alpha}_{-m}^i] =m  $ and   $ (\hat{\alpha}_m^i)^\dagger =  \hat{\alpha}_{-m}^i
$.  The number operator $\hat{N}_m$ for $  m \; \ge 1$ is given by  the product of the annihilation and creation operators as 
$ 
\hat{N}_m^i =  (\hat{\alpha}_{-m}^i  \hat{\alpha}_m^i )  
$. 
Here,  $|k\rangle$ is defined as  $\hat{\alpha}_{-k}^i|0\rangle $, and so  the eigen-states of the number operator satisfy
$ 
\hat{N}_m^i |k\rangle  = k_m^i |k\rangle
$, 
with  $k_m^i $ being the eigenvalue of $\hat{N}_m^i $. 
We can define  a coherent state for world-sheet modes using these operators as
$\hat\alpha^i_m |\varphi^i_m\rangle = \varphi^i_m|\varphi^i_m\rangle  $.
In order to derive an explicit expression for  world-sheet coherent state $ |\varphi^i_m\rangle$,   we expand it in terms of  $ |k\rangle$ string states as 
\begin{equation}
\label{STS8}
|\varphi^i_m\rangle =  \sum_{k=0}^{\infty} |k\rangle \langle k|\varphi^i_m\rangle
\end{equation}
Now we can have an explicit expression for string  world-sheet mode coherent state as 
\begin{equation}\label{60}
|\varphi^i_m\rangle =\sum_{k=0}^{\infty}  \frac{(\varphi^i_m)^{\frac{k}{m}} e^{{-\frac{|{\varphi^i_m}|^2}{2}}} }{\sqrt{k(k-m)(k-2m)...(m)}}  |k\rangle
\end{equation} 
This can then be used to a coherent state for the string in target space as
$ |\Psi\rangle  = \prod_{m\geq1}\prod_{i=1}^{24} |\varphi^i_m\rangle $  \cite{phys}. 

To derive the quantum speed limit, we will have to first obtain a definition of time on world-sheet, which will contain
Fisher information. This can be done using the operator $\hat{M}^2$, which can be defined in terms of the string oscillatory
modes as 
 \begin{equation}
\hat{M}^2 = \frac{1}{\alpha'}\Bigg(\sum_{i=1}^{24}\sum_{n>0}
\hat{\alpha}_{-n}^i\hat{\alpha}^i_n-1\Bigg)
 \end{equation}
 Now $\theta$  is defined as the unobservable variable probed by this operator   \cite{mass, mass1}. 
Since $\theta$ acts as the time for the systems, and $\hat{M}^2$ plays the same role as the Hamiltonian \cite{mass, mass1}, we can write the $\theta$ evolution of a world-sheet mode  $|\varphi^i_{m_0}\rangle$ from $\theta =0$  to $\theta$ as 
 \begin{equation}
  |\varphi^i_{m \theta}\rangle =e^{(-i\hat{M}^2\theta)}|{\varphi^i_{m_0}}\rangle   
 \end{equation}
 Here $\hat{M}^2$ is the generator of $\theta$ translation.  
 Using the relation between world-sheet modes and quantum state of a string in target space \cite{phys}, we can write the  evolution of a  string state  as
  \begin{eqnarray}
  |\Psi_{\theta}\rangle&=&e^{(-i\hat{M}^2\theta)}|\Psi_0\rangle   
  \nonumber \\ &=& e^{(-i\hat{M}^2\theta)}\prod_{i}\prod_{m} \sum_{k}\alpha_m^i|k_0\rangle
 \end{eqnarray}
 Next, we write the Wootters distance for  string theoretical states, which is a statistical  distance $L_{sd}  $,    given by   
 $L_{sd} = {\arccos}\big(|\langle\Psi_{0} |\Psi_{\theta}\rangle)$ \cite{o2, o1}. 
 Next, we observe that $- d   |\langle \Psi_{0} |\Psi_{\theta}  \rangle| /d\theta \leq  | d  \langle \Psi_{0} |\Psi_{\theta}  \rangle /d\theta  |$,  and use the  Schwarz inequality to obtain \cite{okay}  
\begin{equation}
\frac{d L_{sd} }{d\theta}\leq\Big|\frac{d}{d\theta}|\langle\Psi_{0}|\hat{M}^2|\Psi_{{\theta}}\rangle|\Big|
\leq  |\langle\Psi_{0}|\hat{M}^2|\Psi_{{\theta}}\rangle| 
\leq  |\langle\Psi_{ 0}|\hat{M}^2 |\Psi_{{0}}\rangle| 
\equiv  |\langle \hat{M}^2 \rangle|
\end{equation}
Integrating this equation yields a minimal value for  $\theta$ as 
\begin{equation}\label{pp}
    \theta \geq \frac{\pi }{2|\langle \hat{M}^2\rangle|}
    \end{equation}
According to this equation, $\theta$ parameterizes the evolution of   a  string state  to  another orthogonal  string state. This is expected as the unobservable parameter $\theta$ is viewed as time in string theory \cite{mass, mass1}. However, there is a bound on $\theta$, and it is not possible for orthogonal string states to evolve into each other faster than this bound. As this bound is expressed in terms of the average of $\hat{M}^2$ (calculated at $|\Psi_0\rangle$),  it is the string equivalent of Margolus–Levitin bound \cite{7}.  

\begin{figure}[h]
	\centering
	\subfigure[]{\includegraphics[width=0.45\linewidth,height=0.35\linewidth]{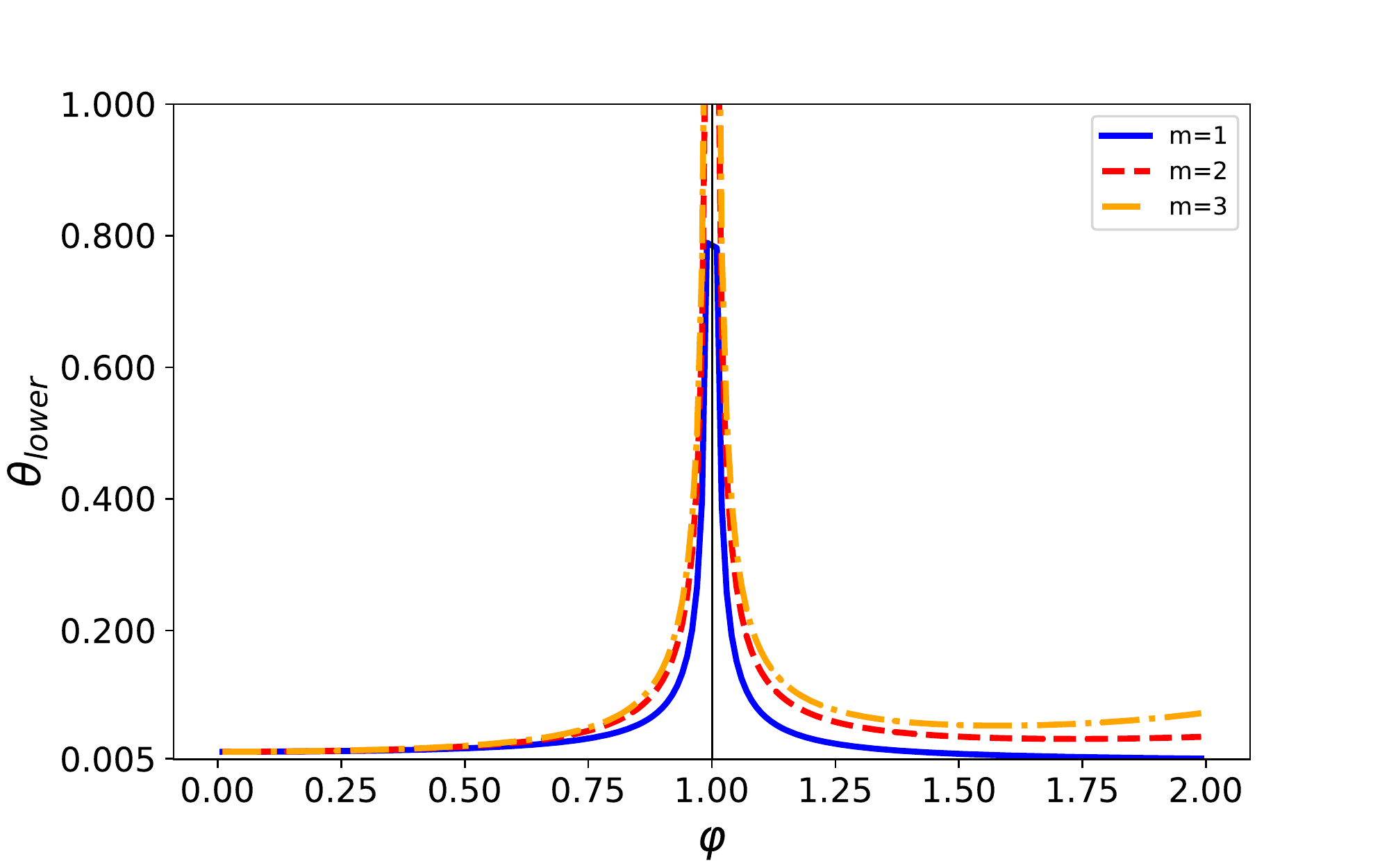}}
	\subfigure[]{\includegraphics[width=0.45\linewidth,height=0.35\linewidth]{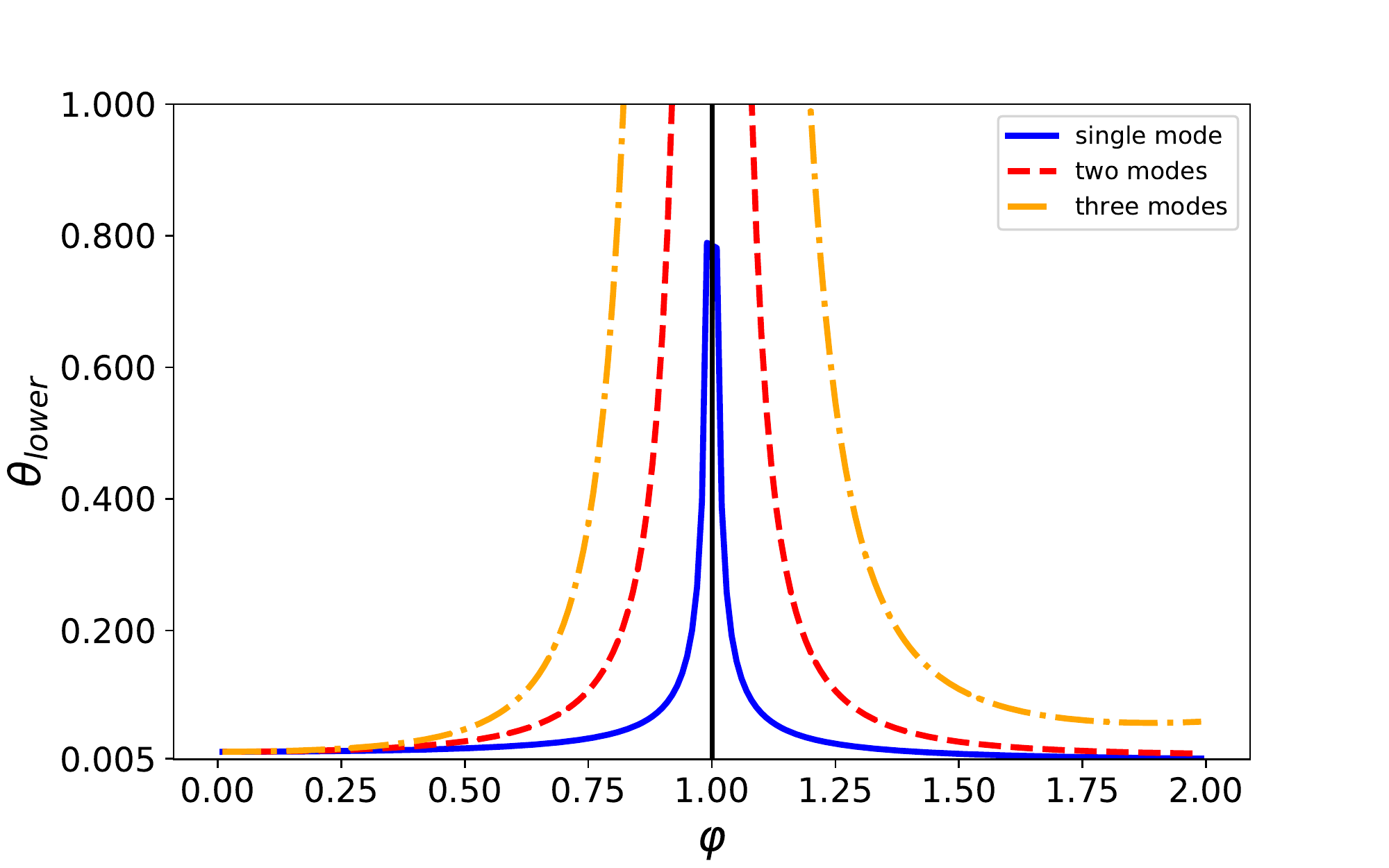}}
	\caption{Plots for the quantum speed limit for (a) different single-mode states and (b) multi-mode states.}\label{fig1}
	\centering
\end{figure}

As we will be calculating every expression at $\theta =0$, we will now drop the $\theta$ index and write $|\Psi_0\rangle$ as $|\Psi\rangle$. 
We can write  $\langle \hat{M}^2\rangle = \langle \Psi|\hat{M}^2|\Psi\rangle$, and express string coherent state  $|\Psi\rangle$   in terms of    $|k\rangle$ states as \cite{phys}
\begin{eqnarray}
\langle \Psi|\hat{M}^2|\Psi\rangle &=& \prod_{i=1}^{24}\prod_{m\geq1}\langle \varphi_m^i|\frac{1}{\alpha^{\prime}}(\hat{N}_m^i -1)|\varphi_m^i\rangle\nonumber\\
&=&\prod_{i=1}^{24}\prod_{m\geq1}\sum_{k}\Big{\langle} k \Big{|}  \frac{(\varphi^i_m)^{\frac{k}{m}} e^{{-\frac{|{\varphi^i_m}|^2}{2}}} }{\sqrt{k(k-m)(k-2m)...(m)}}\Big|  \frac{1}{\alpha^{\prime}}(\hat{N}_m^i-1)\Big|\frac{(\varphi^i_m)^{\frac{k}{m}} e^{{-\frac{|{\varphi^i_m}|^2}{2}}} }{\sqrt{k(k-m)(k-2m)...(m)}}\Big| k\Big{\rangle}\nonumber\\
 &=&   \prod_{i=1}^{24}\prod_{m\geq1}\sum_{k}\frac{(\varphi^i_m)^{\frac{2k}{m}} e^{{-|{\varphi^i_m}|^2}}}{{k(k-m)(k-2m)...(m)}}\Big\{\frac{1}{\alpha^{\prime}}(k-1)\Big{\}}.  
\end{eqnarray}
Using this expression  Eq.\eqref{pp} becomes
\begin{equation}\label{eq17}
    \theta \geq \prod_{i=1}^{24}\prod_{m\geq1}\sum_{k}{\pi }{\Bigg[2\Bigg{|}  \frac{(\varphi^i_m)^{\frac{2k}{m}} e^{{-|{\varphi^i_m}|^2}}}{{k(k-m)(k-2m)...(m)}}\Big{\{} \frac{1}{\alpha^{\prime}}(k-1)\Big{\}}\Bigg{|}}\Bigg]^{-1}.
    \end{equation}
This is a closed form for the  string  Margolus-Levitin inequality. In Fig.\,\ref{fig1}, the lower limit of $\theta$ ($\theta_{lower}$) from Eq.(\ref{eq17}) is plotted against $\varphi$. A singularity is observed at $\varphi=1$ as $\langle M^2\rangle=0$ at this point for single and multi-mode string coherent states. Also, it is clear from the plot (Fig.\,\ref{fig1}\,(a)) that as $m$ increases, $\theta_{lower}$ rapidly approaches a large value for $\varphi<1$ and drops slowly for $\varphi>1$. Similar findings arise when we consider states with an increasing number of modes (Fig.\,\ref{fig1}\,(b)). 
It can be seen from these plots that $\theta \neq 0$, and thus there is a minimal time needed for one state to evolve to another orthogonal state. From \ref{fig1},  we observe that  the minimum value of $\theta$ is $\theta_{lower}\geq 0.005$. 

The Fisher information has been used to define  time on world-sheet \cite{mass, mass1}.
We can write the  Fisher information associated with $\theta$ in terms of the statistical distance, which can then be expressed in terms of $\Delta \hat{M}^2$ as \cite{okay}
\begin{eqnarray}\label{qa}
    F(\theta)= \Big(\frac{dL_{sd}  }{d\theta}\Big)^2 
    &\leq& \frac{\pi}{2}\Big\langle\big(\Delta \hat{M}^2\big)^2\Big\rangle.
\end{eqnarray}
Integrating this expression, we obtain
\begin{equation}\label{qq}
    \theta\geq \frac{\pi}{2} \frac{1}{\big|\sqrt{\langle(\Delta \hat{M}^2)^2\rangle}\big|}.
\end{equation}
This is the string theoretical version of the Mandelstam-Tamm inequality \cite{6}. 
The variance using $  | \langle \Delta{(\hat{M}^2)}^2 \rangle | = | \langle \hat{M}^4\rangle - {\langle \hat{M}^2\rangle}^2 |$. 
Now we observe $ \langle \hat{M}^4 \rangle = \langle \Psi |\hat{M}^4|\Psi \rangle $, and     write  
      \begin{eqnarray}\langle \hat{M}^4\rangle  &=&
      \prod_{m\geq1} \prod_{i=1}^{24}\sum_{k}\Big{\langle} k \Big{|}  \frac{(\varphi^i_m)^{\frac{k}{m}} e^{{-\frac{|{\varphi^i_m}|^2}{2}}} }{\sqrt{k(k-m)(k-2m)...(m)}}\nonumber \\ &&  \times \Big|  \Big(\frac{1}{\alpha^{\prime}}\Big)^2 \big(  \hat{\alpha}_{-m}^i\hat{\alpha}_{-m}^i\hat{\alpha}_m^i\hat{\alpha}_m^i -2\hat{\alpha}_{-m}^i \hat{\alpha}_m^i +1 \big)\Big|
      \frac{(\varphi^i_m)^{\frac{k}{m}} e^{{-\frac{|{\varphi^i_m}|^2}{2}}} }{\sqrt{k(k-m)(k-2m)...(m)}}\Big| k\Big{\rangle}\nonumber \\
       &=&  \prod_{m\geq1} \prod_{i=1}^{24}\sum_{k}\Bigg\{  \frac{(\varphi^i_m)^{\frac{2k}{m}} e^{{-|{\varphi^i_m}|^2}}}{{k(k-m)(k-2m)...(m)}}\Bigg\}\Big(\frac{1}{\alpha^{\prime}}\Big)^2 \Big\{  k(k-m) -2k +1 \Big\}
\end{eqnarray}
Similarly, we can  write 
\begin{eqnarray}\label{nn}
    {\Big\langle \hat{M}^2\Big\rangle}^2
   & =& \Bigg\{\prod_{i=1}^{24}\prod_{m\geq1}\sum_{k}\frac{(\varphi^i_m)^{\frac{2k}{m}} e^{{-|{\varphi^i_m}|^2}}}{{k(k-m)(k-2m)...(m)}}\Big(\frac{1}{\alpha^{\prime}}(k-1)\Big)\Bigg\}^2\nonumber\\
&=& \prod_{i=1}^{24}\prod_{m\geq1}\sum_{k} \prod_{j=1}^{24}\prod_{n\geq1}\sum_{k^\prime}\Bigg\{ \frac{(\varphi^i_m)^{\frac{2k}{m}}(\varphi^j_n)^{\frac{2k^\prime}{n}} e^{{-|{\varphi^i_m}|^2}} e^{{-|{\varphi^j_n}|^2}}}{{\{k(k-m)...(m)\} .\{ k^\prime(k^\prime-n)...(n)\}}}\Big(\frac{1}{\alpha^{\prime}}\Big)^2 (k-1)(k^\prime -1)\Bigg\}
\end{eqnarray}
Now using these equations, we obtain the variance of $\hat{M}^2$. Using Eq.(\ref{qq}), we can write the bound on $\theta$ as 
\begin{eqnarray}
    \theta &\geq& \frac{\pi  }{2}  \prod_{i=1}^{24}\prod_{m\geq1}\sum_{k}\Bigg| \Bigg[\Bigg\{  \frac{(\varphi^i_m)^{\frac{2k}{m}} e^{{-|{\varphi^i_m}|^2}}}{{k(k-m)(k-2m)...(m)}}\Bigg\}\Big(\frac{1}{\alpha^{\prime}}\Big)^2 \Big\{  k(k-m) -2k +1 \Big\}\nonumber\\
    &-& \prod_{j=1}^{24}\prod_{n\geq1}\sum_{k^\prime}\Bigg\{ \frac{(\varphi^i_m)^{\frac{2k}{m}}(\varphi^j_n)^{\frac{2k^\prime}{n}} e^{{-|{\varphi^i_m}|^2}} e^{{-|{\varphi^j_n}|^2}}}{{\{k(k-m)...(m)\} .\{ k^\prime(k^\prime-n)...(n)\}}}\Big(\frac{1}{\alpha^{\prime}}\Big)^2 (k-1)(k^\prime -1)\Bigg\} \Bigg]^{-\frac{1}{2}} \Bigg|
\end{eqnarray}
\begin{figure}[t]
	\centering
	\subfigure[]{\includegraphics[width=0.45\linewidth,height=0.35\linewidth]{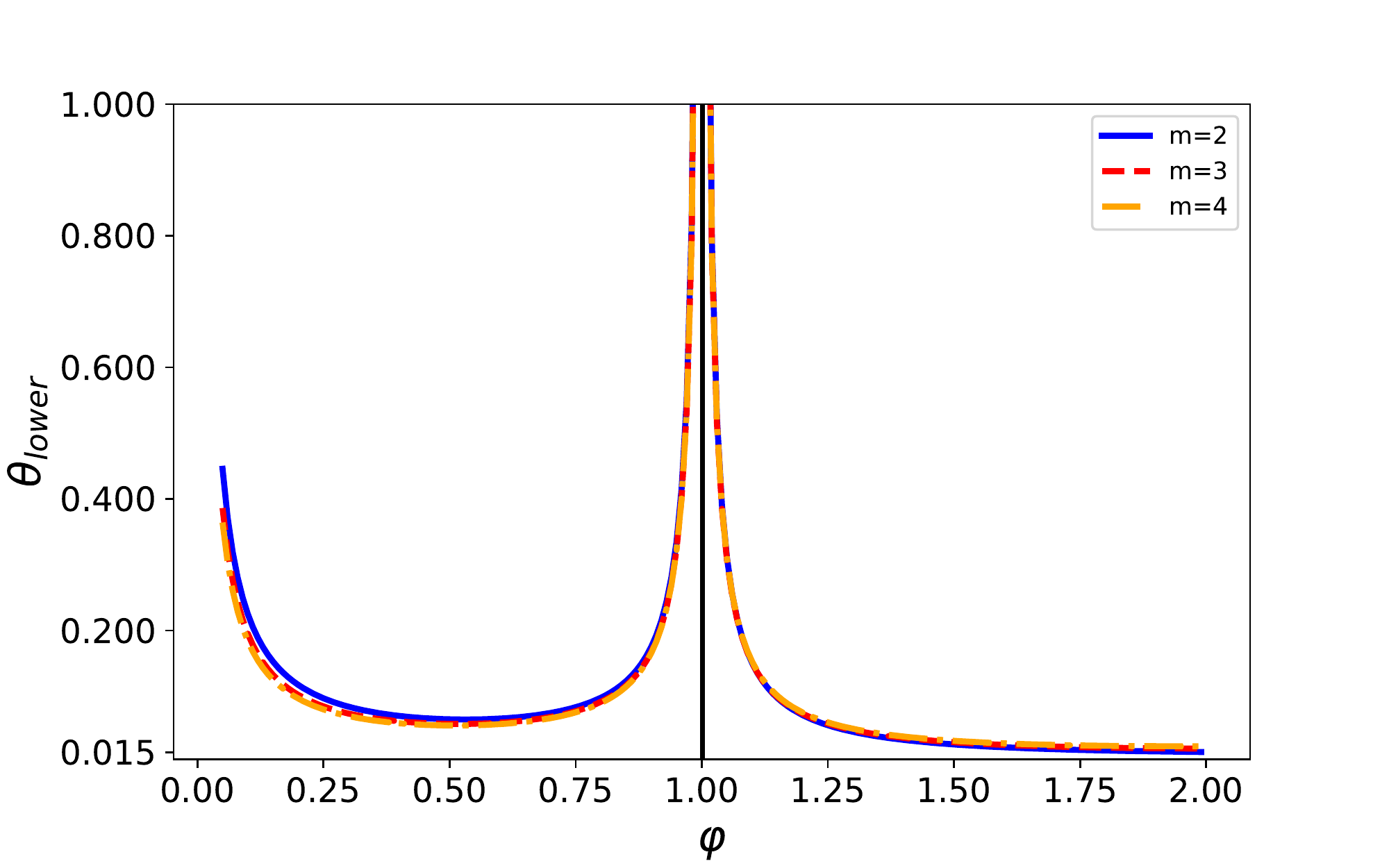}}
	\subfigure[]{\includegraphics[width=0.45\linewidth,height=0.35\linewidth]{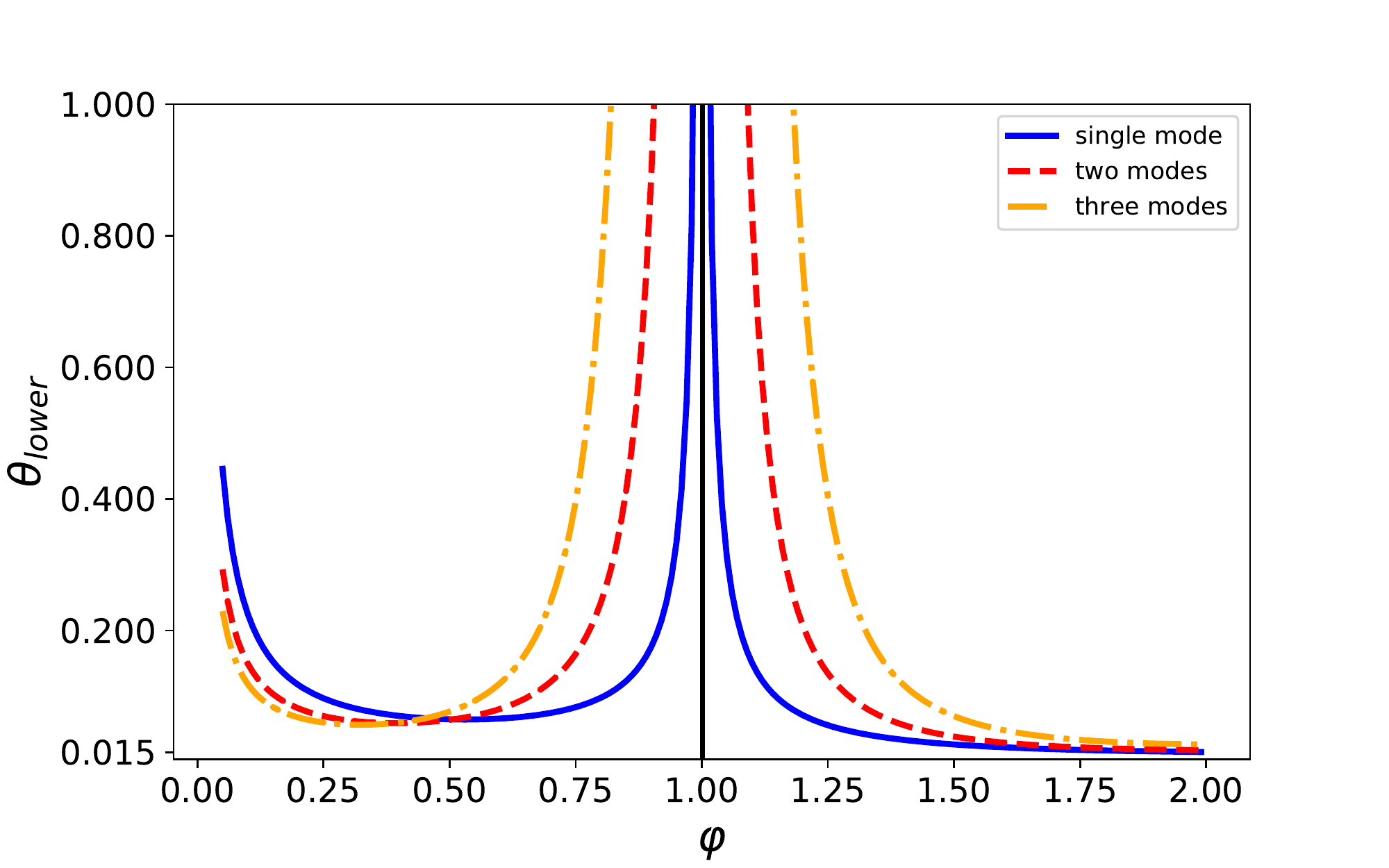}}
	\caption{Plots for $\theta_{lower}$ vs $\varphi$ for (a) different single-mode states and (b) multi-mode states.}\label{fig2}
	\centering
\end{figure}
This is the explicit form of string theoretical Mandelstam-Tamm inequality. 
To have an insight, we plotted in Fig.\,\ref{fig2} the lower limit of $\theta$ vs. $\varphi$ of the above equation. Unlike the
single-mode scenario in Eq.(\ref{eq17}), here the quantum speed limit is the same for all $m$ values (Fig.\,\ref{fig2}\,(a)).
However, as the number of modes increases, the $m4$ values vary in a similar pattern to Fig.\,\ref{fig1}\,(b) and is
given in Fig.\,\ref{fig2}\,(b). From \ref{fig2},  we observe that  the minimum value of $\theta$ is $\theta_{lower}\geq 0.015$.

The interactions in quantum field theory represent certain particles becoming different kinds of particles, which are
effective descriptions of a fundamental string changing its modes. However, as there is minimal time required
for any such evolution to occur, the interactions will need a minimal time $\theta$. In this letter, we observed that such minimal time $\theta_{lower}\neq 0$ occurs using the quantum speed limit in string theory. 
Thus, the interactions will be smeared
over this interval $\theta_{lower}$.  For any effective quantum field theory with interactions $f(\phi)$, involving an
effective field $\phi$, we have to replace the local  interaction with   $h(f(\phi), \theta_{lower})$, where $h$
is a non-local function obtained by smearing $f(\phi)$ over the interval $\theta_{lower}$.  We note that it is
already known that such non-locality occurs in effective field theory obtained from  string theory \cite{st12, st14}. 
The star product of the second quantized string theory smears out the interaction over the string length producing non-locality \cite{sf12, sf14}.  However,  using field redefined,  the interaction can have the usual polynomial
form, and the non-locality will be absorbed in the kinetic part of the action \cite{sf15, sf19}. Thus, it is expected
that all quantum field theories should be replaced by non-local quantum field theories due to the quantum speed limit in
string theory.  Even though it is known that string
theory will produce non-locality \cite{st12, st14}, we have argued that the actual theoretical reason for it is related
to the quantum speed limit in string theory. 

The non-locality field theories can have important physical consequences. It is known that string theory can be approximated by general relativity coupled to other fields in higher dimensions \cite{general}. However,   this is only an approximation, obtained by neglecting some important effects.     Thus, as pointed out in this letter, quantum speed limit in string theory  should produce non-locality in quantum field theories. 
General relativity can be studied  using the formalism of perturbative quantum field theories, and this is done in perturbative quantum gravity \cite{pert1, pert2}. However, perturbative quantum gravity is a non-renormalizable theory \cite{2b, 3b}. 
The results of this paper suggest that due to the quantum speed limit, general relativity should be replaced by non-local gravitational action. Even though we have not obtained this gravity action explicitly, the existence of non-local gravity has been argued on general grounds. 
Similarly, it is also known that divergences do not occur in non-local quantum field theories \cite{a1, a2, a3, a4, a5}. 
Thus, such a non-local modification to general relativity from the results of this letter can resolve the problem of non-renormalizability in perturbative quantum gravity \cite{2b, 3b}.  It has been proposed that non-local modifications can produce a positive cosmological constant \cite{cc1, cc2}. However, it is hard to construct de Sitter solutions with a positive cosmological constant in string theory using the usual approximation  of string theory to general relativity \cite{cc4}. In fact, it has been argued that it might not be possible to construct de Sitter vacua in string theory using general relativity coupled to other fields as a low energy effective field theory    \cite{cc5}.  It would thus be interesting to analyze non-local modifications to the actions obtained as low energy effective field theories from string theory and analyze the construction of de Sitter in such theories. It is expected that such non-locality will produce a cosmological constant, as the cosmological constant in quantum field theories can be produced from non-locality \cite{cc1, cc2}. It would be interesting to investigate the construction of such non-local actions, and then analyze the construction of de Sitter spacetime using them. 


In this work, we used a novel approach to address the problem of time on world-sheet \cite{mass, mass1}. 
It has been argued that time in string theory should be seen as  an unobservable variable,  which one may only indirectly
probe  through $\hat{M}^2$ \cite{mass, mass1}.  This definition of time has been used to obtain the  quantum speed
limit in string theory. We   first used $\hat{M}^2$ to probe time, and then calculated   both the Mandelstam-Tamm bound and
Margolus–Levitin bound for string theory to obtain   a minimum time for a particle state evolving into another particle state.
This has a novel  consequence for  divergences in quantum field theories as minimum time implies  that
divergences should not occur in quantum field theories. Furthermore, it has been argued that the cutoff actually corresponds
to indirectly imposing such a quantum speed limit. We have also discussed the physical applications of these results.


\begin{thebibliography}{99}
\bibitem{1b}  S-k. Ma, {Reviews of Modern Physics},    {45},  (1973), 589.
\bibitem{2b} G. ’t Hooft and M. Veltman, Ann. Inst. H. Poincare Phys. Theor. A    {20},  (1974), 69.
\bibitem{3b} S. Deser and P. van Nieuwenhuizen, Phys. Rev. D    {10},  (1974), 401.
\bibitem{6} L. Mandelstam and I. G. Tamm, J. Phys. (USSR)    {9},  (1945), 249.
\bibitem{4b} S. Lloyd,  
Nature      {406},   (2000), 1050.
\bibitem{a1} J.~W.~Moffat, Phys. Rev. D     {41},  (1990), 1180.
\bibitem{a2} D.~Evens, J.~W.~Moffat, G.~Kleppe and R.~P.~Woodard, Phys. Rev. D     {43},  (1991), 510.
\bibitem{a3} L.~Modesto and L.~Rachwal,
 Int. J. Mod. Phys. D     {26},  (2017), 1730020.
\bibitem{a4} G.~Calcagni and L.~Modesto, Phys. Rev. D     {91},   (2015), 124059.
\bibitem{a5} S.~Denk, V.~Putz, M.~Schweda and M.~Wohlgenannt, Eur. Phys. J. C     {35},  (2004), 287.
\bibitem{7} N. Margolus and L. B. Levitin, Physica D    {120},  (1998), 188. 
\bibitem{16}  J. Anandan and Y. Aharonov,  Phys. Rev. Lett.    {65},  (1990), 1697.
\bibitem{17} P. J. Jones and P. Kok,  Phys. Rev. A    {82},  (2010), 022107.
\bibitem{18} M. Zwierz, Phys. Rev. A    {86},  (2012), 016101.
\bibitem{19}  B. Russell and S. Stepney, arXiv:1410.3209.
\bibitem{20} D. P. Pires, M. Cianciaruso, L. C. C´eleri, G. Adesso,
and D. O. Soares-Pinto,  Phys. Rev. X    {6},  (2016), 021031.
\bibitem{10}A. Uhlmann,  Phys. Lett.
A    {161},  (1992), 329.
\bibitem{11} P. Pfeifer,  Phys. Rev. Lett.    {70},  (1993), 3365.
\bibitem{12} S. Deffner and E. Lutz,  J. Phys. A    {46},  (2013), 335302.
\bibitem{13}  M. M. Taddei, B. M. Escher, L. Davidovich, and R. L.
de Matos Filho,  Phys. Rev. Lett.    {110},  (2013), 050402.
\bibitem{14}A. del Campo, I. L. Egusquiza, M. B. Plenio, and S. F.
Huelga,  Phys. Rev. Lett.    {110},  (2013), 050403.
\bibitem{15} S. Deffner and E. Lutz,  Phys. Rev. Lett.    {111}, 
(2013), 010402.

\bibitem{21}  K. Bhattacharyya, J. Phys. A    {16},  (1983), 2993.
\bibitem{22} S. Luo,  Phys. D    {189},  (2004), 1.
\bibitem{23}L. B. Levitin and T. Toffoli,  Phys. Rev. Lett.    {103},  (2009), 160502.
\bibitem{24}T. Caneva, M. Murphy, T. Calarco, R. Fazio, S. Montangero, V. Giovannetti, and G. E. Santoro, Phys. Rev. Lett.
   {103},  (2009), 240501.
\bibitem{25}S. Deffner and E. Lutz,  Phys. Rev. Lett.    {105}, (2010), 170402.
\bibitem{26} M. Murphy, S. Montangero, V. Giovannetti, and T.
Calarco,  Phys. Rev. A    {82},  (2010). 022318.
\bibitem{27} G. C. Hegerfeldt,  Phys. Rev. Lett.
   {111},  (2013), 260501.
\bibitem{28} D. Mondal and A. K. Pati,  Phys.
Lett. A    {380},  (2016), 1395.
\bibitem{time12}W. Pauli, General Principle of Quantum Theory Springer, Berlin (1980)
 
   \bibitem{eq1}S. L. Braunstein and C. M. Caves,  Phys. Rev. Lett.    {72},  (1994), 3439.
  \bibitem{eq2} G. Toth and D. Petz
Phys. Rev. A    {87},  (2013), 032324.
\bibitem{eq4}M.  G. A. Paris, 	Int. J. Quant. Inf.    {7},  (2009) , 125.
\bibitem{eq5} M. G. Genoni, P. Giorda and M. G. A. Paris, Phys. Rev. A    {78},  (2008), 032303.
\bibitem{eq6}G. Brida, I. Degiovanni, A. Florio, M. Genovese, P. Giorda, A. Meda, M. G. A. Paris and A. Shurupov, Phys. Rev. Lett.    {104},  (2010), 100501.
\bibitem{eq7}  V. D'Auria, S. Fornaro, A. Porzio, S. Solimeno, S. Olivares and   M. G. A. Paris, Phys. Rev. Lett    {102},  (2009), 020502.


\bibitem{mass}S.~S.~Wani, A.~Shabir, M.~Faizal and S.~Rubab, EPL     {139},   (2022),42002.
\bibitem{mass1} S.~S.~Wani,  J. Q. Quach and  M.~Faizal,  EPL    {139},  (2022), 62002.

\bibitem{t12} S.~S.~Wani, J.~Q.~Quach, M.~Faizal, S.~Bahamonde and B.~Pourhassan,
 Found. Phys.    {52}  (2022), 25

\bibitem{pt12}E. Anderson,
 Annalen Phys.  {524}  (2012), 757
\bibitem{pt14} R.  M. Wald, 
Phys. Rev. D 48 (1993),  R2377(R)
  \bibitem{wh}S. Carlip, Phys. Rev. Lett. 123 (2019), 131302
\bibitem{wh0}J.~J.~Halliwell, Phys. Rev. D {38} (1988), 2468
 \bibitem{wh2}M. Bojowald, M. Kagan, H. H. Hernández and A. Skirzewski, Phys. Rev. D 75 (2007), 064022 
 \bibitem{wh21} L. Smolin, 
Phys. Rev. D 84 (2011), 044047
 \bibitem{wh4}H. W. Hamber and R. M. Williams, 
Phys. Rev. D 84 (2011), 104033
\bibitem{wh41}B. Baytas and M. Bojowald, 
Phys. Rev. D 95 (2017), 086007
 \bibitem{wh5} A. Baratin and D. Oriti, 
Phys. Rev. Lett. 105 (2010), 221302
 \bibitem{wh51}S. Gielen, D. Oriti and L. Sindoni, 
Phys. Rev. Lett. 111 (2013), 031301
 \bibitem{wh6}X. Zhang and Y. Ma, 
Phys. Rev. Lett. 106 (2011), 171301 
 \bibitem{wh61}A. Alonso-Serrano, M. Bouhmadi-Lopez and P. Martin-Moruno,  Phys. Rev. D 98 (2018), 104004
 \bibitem{wh7}A. Boyarsky, A. Neronov and I. Tkachev, 
Phys. Rev. Lett. 95 (2005), 091301
\bibitem{wh71} P. Gusin, 
Phys. Rev. D 77 (2008), 066017 
\bibitem{wh8}E.~Rodrigo, Phys. Lett. B  160 (1985), 43
\bibitem{wh81} E.~Rodrigo, Phys. Lett. A  {105} (1984), 196
 \bibitem{qw} S. Deffner, S. Campbell, J. Phys. A,    {50}, (2017), 453001.
 \bibitem{phys}S.~S.~Wani, A.~Shabir, J.~U.~Hassan, S.~Kannan, H.~Patel, C.~Sudheesh and M.~Faizal,
Annals of Phys. {441} (2022), 168867
\bibitem{o2} S. L. Braunstein and C. M. Caves, Phys. Rev. Lett. 72
(1994), 3439
\bibitem{o1}W. K. Wootters, Phys. Rev. D 23 (1981), 357
\bibitem{okay} P. J. Philip and P. Kok,
  Phys. Rev. A {82} (2010), {022107}
 \bibitem{st12}A.~Belenchia, D.~M.~T.~Benincasa, S.~Liberati, F.~Marin, F.~Marino and A.~Ortolan,
Phys. Rev. Lett. {116} (2016), 161303
 \bibitem{st14}T.~Biswas, J.~Kapusta and A.~Reddy,
JHEP  {12} (2012), 008
\bibitem{sf12}  L. Joukovskaya, Phys. Rev. D 76  (2007), 105007

\bibitem{sf14}T. Biswas, J. A. R. Cembranos and J. I. Kapusta, JHEP 1010  (2010), 048
\bibitem{sf15}V. A. Kostelecky and S. Samuel, Phys. Lett. B 207 (1988), 169
\bibitem{sf19} V. A. Kostelecky and S. Samuel, Nucl. Phys. B 336 (1990), 263
\bibitem{general} S. Mukhi, Class. Quant. Grav. 28 (2011), 153001

\bibitem{pert1}A.~Eichhorn, S.~Lippoldt, J.~M.~Pawlowski, M.~Reichert and M.~Schiffer,
Phys. Lett. B  {792} (2019), 310
\bibitem{pert2} R.~Akhoury, R.~Saotome and G.~Sterman, Phys. Rev. D  {103} (2021), 064036

\bibitem{cc1}I.~Oda, Eur. Phys. J. C  {78} (2018), 294
\bibitem{cc2}I.~Oda, Phys. Rev. D  {96} (2017), 024027
\bibitem{cc4}M.~Dine, J.~A.~P.~Law-Smith, S.~Sun, D.~Wood and Y.~Yu,
JHEP  {02} (2021), 050
\bibitem{cc5} U.~H.~Danielsson and T.~Van Riet, Int. J. Mod. Phys. D  {27}   (2018), 1830007

 



%



 
 

\end{thebibliography}
\end{document}